# REVIEWING METI: A CRITICAL ANALYSIS OF THE ARGUMENTS

**JOHN GERTZ***
*Zorro Productions, 125 University Avenue, Suite 101, Berkeley, CA 94710, USA.*
Email: jgertz@firsst.org

There is an ongoing debate pertaining to the question of whether Earth should initiate intentional and powerful radio transmissions to putative extra-terrestrial (ET) civilizations in the hope of attracting ET's attention. This practice is known as METI (Messaging to ET Intelligence) or Active SETI. The debate has recently taken on a sense of urgency, as additional proponents have announced their intention to commence *de novo* transmissions as soon as they become funded and acquire the needed time on a powerful transmitter such as Arecibo. Arguments in favor of METI are reviewed. It is concluded that METI is unwise, unscientific, potentially catastrophic, and unethical.

## 1. INTRODUCTION

In the medical sciences, proposed experiments must pass ethics review boards. Some experiments are simply too dangerous or unethical to be performed, certainly not just on one's own lonely say-so. We do not clone humans; we do not conduct table top experiments with smallpox; and we no longer inject human subjects with pathogens in order to trace the course of a disease or to see how long it might take for subjects to die. Though a commonplace in medical research, astronomers face no such ethical reviews, since theirs is normally an observational science only. When it comes to METI (Messaging to ET Intelligence, also called or Active SETI), which is not observational but manipulative, and on which may hinge the very fate of the world, perhaps they should.

Do space aliens present a clear and present danger and, if so, is there anything we can do about it? There is not one scintilla of credible evidence that Earth has ever been visited by space aliens, much less that aliens have sought to do damage to the Earth. However, extraterrestrials (ET), if they exist, may soon learn that Earth harbors technologically advancing life forms, and that may change everything. Our electromagnetic (EM) emissions leave Earth at the speed of light. EM that left Earth in 1930 has already swept over approximately the nearest 7,000 stars.

That said, Earth's EM leakage is either very weak, not pointed at nearby stars, or both. Further, the Earth grows quieter annually as more information is transmitted via cable, the Internet, and satellites rather than terrestrially over the air. Unless ET's receivers are both powerful and omnidirectional, they will not detect us. ET's receivers could be omni-directional, but unable to pick up a signal so weak as the proverbial *I Love Lucy*. For example, the gigantic Arecibo radio telescope could not detect terrestrial TV transmissions, if broadcast from the distance of our nearest neighboring stars. Alternatively, an ET receiver could be very powerful, but it might take millennia for it to get around to slewing in our direction, given the large number of potential targets. By the time Earth returns into ET's crosshairs for a routine check in, we might have gone silent.

The first modern SETI search was conducted by Frank Drake in 1960 [1]. From that date until today, there has been no agreed upon detection of an alien signal. Some are now arguing that since so much time has elapsed without success, it is time to announce ourselves to ET by using our most powerful radio telescopes as transmitters in order to proactively send our signals to Earth's nearest stars in an effort to attract ET's attention. Arecibo, for instance, is so powerful that, when used as a transmitter, its signal is potentially capable of being detected at vast interstellar distances.

A new consideration of the METI debate assumes some urgency at this time. When the SETI Institute (SI) rejected a proposal from Vakoch and Shostak to initiate immediate high power radio transmissions directed to Earth's neighboring stars, Vakoch founded another organization, METI International [2, 3], with the same intent [4, 5]. Fearing a gathering storm, a cohort of SETI scientists and thinkers issued a statement in opposition to METI in February, 2015 [6].

The current paper will further consider the arguments of METI's proponents (METI-ists) and opponents.

## 2. METI-IST ARGUMENTS EXPLAINED

### 2.1 After so Many Years of Failure, it is Time to Try a New Approach

METI-ists argue that because we have conducted SETI searches for more than five decades without success it is therefore time to try a different approach. However, SETI searches have barely begun to scratch the surface. For example, the SETI Institute (SI) has only examined less than one star in 50 million in the Milky Way. Even then, this limited set has been studied in real time for only ten minutes each, only across certain frequencies, and only using certain detection algorithms. Jill Tarter, SI's lead SETI scientist for most of its history, often likens this to having dipped a drinking glass into the ocean. The fact that no fish appear in that first dip of the glass hardly means that the ocean is lifeless. However, our SETI searches today are vastly more powerful than those conducted in previous decades by virtue of

* Foundation for Investing in Research on SETI Science and Technology (FIRSST), Berkeley, California.

(a) better equipment designs; (b) more sophisticated algorithms; and (c) most importantly, because the benefits that have accrued from Moore's Law have brought an exponential improvement in computer processing power. Additionally, for most of its history, SETI has not received government funding, and has therefore mostly relied on limited private philanthropy. The funding situation suddenly and dramatically improved when in 2015 Yuri Milner's Breakthrough Foundation announced its pledge of $100 million over ten years to conduct the most powerful (by at least three orders of magnitude) SETI search ever [7]. This has led SETI scientist, Seth Shostak, to predict in many forums that ET will be detected within the next twenty years.

### 2.2 ET May be Waiting for us to Call

If extant, ET is eerily silent. It is a mystery as to why ET's presence is not as obvious to us as, say, the full moon. Extrapolating our technology out several centuries, we can imagine that it might take less than 50 million years to colonize the entire galaxy [8]. This presumes only that we have the technology to travel to and colonize planets around our nearest stars. After a few centuries of domestic development, those colonies might then send forth new spacecraft to colonize yet further stars. The empire spawned by Earth in this fashion would grow exponentially in all directions. So where are all the empires started by other civilizations [9]?

There have been dozens of possible explanations put forth to explain this so-called Great Silence, or Fermi's Paradox [10]. Perhaps mankind really is all alone. Perhaps our search strategies are wrong. We currently search for radio and optical transmissions, whereas ET might prefer to send physical, information laden probes to orbit our sun. No serious search for such probes has yet been conducted. The idea favored by METI-ists is one or another variant of the Zoo Hypothesis, namely, that ET regards us as a nature preserve, and that the galactic rule is that newcomers are to be left alone until they signal their wish to join the galactic club. To give METI-ists their due, this is a perfectly plausible possibility. Alternatively, maybe the galaxy is so silent because ET knows something that we do not, namely, that real planet killing danger lurks out there [11].

### 2.3 The Cat is Out of the Bag Anyway

METI-ists argue that ET can surely monitor Earth already. They speculate that, be it with a gravity lens telescope, or by means of a terrestrial radio telescope of vast dimensions, ET can detect our leakage. For example, METI-ists have noted that a gravity lens telescope orbiting ET's home star at 550 AU should be capable of detecting Earth's EM leakage if ET's home star is reasonably close to ours [12, 13]. Let's grant the point. The problem is that such a telescope would have a very fine focal point centered precisely opposite ET's home star. Earth would come into that focus only once during each of the telescope's 13,000 year orbits, given a sun-like host star, and then only if Earth just happened to be on the very narrow ribbon that orbit traced out. The focal ribbon would only be about 3-4 arc seconds in width. Because Earth would remain within the crosshairs of the telescope for a short amount of time very little could be learned about the leakage. Therefore, ET would have to deploy many millions of these gravity telescopes in order to reliably detect Earth's leakage of the last 100 years. The ineffectiveness of ET's conventional radio telescopes in detecting our leakage has been analyzed by Billingham & Bedford, who conclude that ET's radio telescopes would not only have to be truly gargantuan, but pointed at Earth for an extremely long period of time in order to detect our leakage. [14, 15, 16].

Were it in fact true that Earth's leakage has been detected, would it not then follow that there would be no further point to METI? If ET already has detected our transmissions, if they have already tuned in to *I Love Lucy*, we would already have sent our message, and hopefully given ET a good laugh in the bargain. ET may have already sent us their best sitcom in response. Should we not therefore double and triple our SETI efforts in search of that response, rather than waste time and money on METI? METI-ists are entirely disingenuous, since they propose to use Arecibo, the world's most powerful transmitter, which is some $10^5$ times more powerful than the omni-directional leakage they claim ET can already detect. Whereas Earth's EM leakage whispers into the universe, they propose to shout [8, 4]. Conceding this, METI-ists then argue that only Arecibo will suffice because it will be Arecibo that will be used to transmit a return message in the event that an artificial signal is detected and the Earth decides to respond. They have to be prepared and practiced in the use of Arecibo as a transmitter, since in the aftermath of a detection and with little or no notice, they may be called upon to use Arecibo to transmit humanity's officially sanctioned response. But then why propose, as they do, to transmit to the nearest stars? If their argument is sincere, why not transmit only in the direction of a faraway galaxy that is more or less perpendicular to the plane of the Milky Way and which is not occulted by any foreground stars in our galaxy?

#### *2.3.1 Since Arecibo Already Tracks Asteroids when used as a Powerful Radar why not Transmit Intentional Messages?*

When used as a transmitter, usually to track asteroids, Arecibo becomes the most powerful radar on Earth. Goldstone and Evpatoria have also been used as powerful radar transmitters. Their beams can potentially be detected at vast interstellar distances. Fortunately, Arecibo's beam is very narrow, and so it would be highly unlikely for a nearby star to just happen to lie right behind a given asteroid and be thereby inadvertently illuminated by that beam [13]. Moreover, Arecibo has only been used rarely and for short durations as a powerful radar. Some METI-ists have misleadingly conflated these rare, but powerful, narrow beam transmissions with Earth's omni-present and omi-directional, but very weak emissions to assert that ET must have detected us [17, 18]. The asteroid detection radar problem is very easy to fix by adopting a standard of best practices that includes a provision for muting the radar during moments when the target occults a nearby star or transits the plane of the Milky Way. Although Zaitsev has argued that there have been vastly more of these radar transmissions than METI transmissions [19], this misses the very crucial distinction that whereas these radar are not aimed at nearby stars (except unintentionally), METI transmissions, by definition, are.

#### *2.3.2 We Are Only Following in the Footsteps of Other METI-ists*

There have only been tiny dribs and drabs of METI to date. METI-ists, for example, point to the plaques on the Pioneer and Voyager spacecraft or messages borne by New Horizon. These are slow spacecraft headed on trajectories to nowhere in particular. They were not sent for the purpose of METI. More importantly, their very medium is the message. Any ET who encounters one will instantly know it is artificial, will be able to assess the technology of its creators, and be able to deduce its point of origin from its trajectory. In this light, its explicit message from mankind would be tertiary, in the sense that it probably would do no further harm. In 1974, Frank Drake briefly broadcast an intentionally feeble message from Arecibo to M13, a globular cluster some 29,000 light years (LY) away. There is a big difference between broadcasting to M13 at 29,000 LY and a nearby star that is, say, a mere 29 LY in distance. The latter message would be one million times louder. Drake never intended that his METI transmission would actually be received, but merely sought to demonstrate a proof of concept (he has also told me that he regrets having done it). There have been a few other METI attempts, mostly weak and small, and sometimes conducted by crass marketers [20]. Crucially, to use prior METI attempts as justification for future METI efforts breaks the fundamental rule most of us learn in kindergarten, namely, that two wrongs do not make a right.

### 2.4 ET is All Sweetness and Light, so Let's Join the Galactic Club

METI proponents speculate that ET must be imbued with altruism and cooperation. Otherwise, they would certainly have self-destructed before achieving their high level of technology. Intelligent species (e.g., dolphins) can be highly social within their own species and maybe even benign to some outside species (e.g., humans), but they can also be viewed as vicious and heartless killers from the vantage point of yet other intelligent species (e.g., squid). Moreover, SETI theorists have often speculated that a technologically intelligent carbon based life form would soon evolve itself into silicon AI forms [21]. Would altruism be programmed into such AI silicon beings? Would this first generation of self-replicating AI silicon-beings program altruism into their potentially very rapidly evolving descendants? Those later generation AI beings might be so far advanced relative to human consciousness that we could understand virtually nothing about their motivations. We have absolutely no idea what ET would conclude about our civilization upon detection. METI-ists speculate that ET will receive our intentional communication as a signal that we are ready to join the galactic club. Surely, they will send us a laminated membership card along with a welcoming gift basket, included in which will be our very own embossed copy of Encyclopedia Galactica, filled with great wisdom, science, technology and culture. It might, however, just as plausibly be speculated that ET will receive Earth's uninvited intentional communication with the exclamation, "OMG, they know we are here! Let's snuff them before they snuff us!" After all, if ET is watching our *Nightly News* as well as *I Love Lucy*, they will know what a vicious and wretched species we are.

### 2.5 Even if ET Wanted to, it Could not Harm us From a Vast Interstellar Distance [22]

About the only thing we can say about ET with near certainty is that it is more advanced than us, as operantly defined as possessing the capability to send and receive radio, laser, EM transmissions at other frequencies, or physical probes. This statement comes close to being a Law of SETI. Some statistical statements are so strong that they can be deemed laws. The Second Law of thermodynamics is a well-known example. Simply put, given the billions of years hypothetically available to it, the chances that ET is also in its first century of the technological ability to send and receive EM signals is vanishingly small. Nevertheless, as advanced as they might be, a carbon-based ET will probably not travel hundreds or thousands of LYs just to eat us. Big Macs cannot be that expensive on its home planet. Nor are they likely to spend 50, 500, 5000, or however many of their generations traveling here just to conduct a bombing run. They cannot hate us that much (or so we hope). Nor would they care much about our raw materials. They are not likely to come all the way here for water or minerals, which are no doubt as ubiquitous in their system as in ours. Nonetheless, it is specious to suggest that aliens could not harm us if they wished. The Hitittes would have been incredulous at the suggestion of warfare from a distance of more than the short range of their arrows. They could not imagine missiles, artillery and bombers. Contrary to sci-fi movies, ET would not need a space armada in the style of *Independence Day* to destroy life on Earth. A single bullet sized projectile filled with the right self-replicating pathogen or nano-grey-goo might do the job.

Alternatively, ET might employ a fairly small kinetic projectile accelerated to a significant fraction of the speed of light [8]. The asteroid that did in the dinosaurs was traveling at about 6 miles a second, or a mere 0.003% of the speed of light – a very lazy crawl. Such projectiles could be launched from ET's home systems, just as we have launched Pioneer, Voyager and New Horizon into interstellar space from ours. Albeit, our projectiles are not directed against another planet, are travelling at relatively slow speeds, and are not lethal warheads. Musso [20] asserts that interstellar space flight, be it by populated craft, probe or warhead, is probably not possible since otherwise we would have already seen evidence for this. ET's probes could not exist in our solar system because "there should be some evidence for their presence, while, on the contrary, it completely lacks." This is misleading. We currently have no evidence for or against ET probes orbiting the sun in large part because we have almost never explicitly looked for them, though it must also be conceded that they might have shown up by serendipity during radio or optical explorations of background stars or galaxies if they were actively and persistently broadcasting to Earth.

### 2.6 There Is No Law Against It

#### *2.6.1 We do this Altruistically in the Name of all Mankind*

METI is not science, it is unauthorized diplomacy, and is explicitly forbidden under the so-called First Protocol, adopted by the International Academy of Astronautics in 2000 as a proposal to the UN Committee on the Peaceful Uses of Outer Space. The First Protocol (formally, "Declaration of Principles for Activities Following the Detection of Extraterrestrial Intelligence"), Principle #8 of which states: "No response to a signal or other evidence of extraterrestrial intelligence should be sent until appropriate international consultations have taken place." What applies in the aftermath of detection, must surely apply before [10, 23]. The First Protocol does not presently carry the force of law, but it does represent a consensus statement of best practices [8, 24].

METI-ists presume to speak for all mankind. This is anti-democratic, since they would give none of the rest of us an opportunity to agree to their transmissions, or any control over the content of their message. Sensitive to this criticism, METI-ists have proposed to simply upload the Internet [21]. As a businessman, this author particularly objects to this. The entire academic argument as to whether ET is altruistic or predatory [22] may very well be moot. It is entirely possible that ET neither seeks our destruction nor our salvation, but rather seeks simply to trade. Information might be the most valuable—perhaps the only—currency of trade in the galaxy. Give all of our culture, religion, technology and science away for free and ET might laugh up its sleeve at such fools. Why should they bother to respond? What more would they have to gain, especially since communication might involve significant risk that either the intended recipient or an eavesdropper in our star's foreground or background might be hostile? Would it not make better sense to give ET small samples as a loss leader, say a few opuses from Beethoven, Balinese gamelan, and the Beatles; plus some paintings by Rembrandt and Jackson Pollack; plus the equations of Maxwell, but not yet Einstein. Let ET then barter for the rest.

#### 2.6.2 Free Speech

METI-ists argue that they are simply exercising their free speech rights. Free speech is not absolute and is widely acknowledged to exclude provocations to violence. METI transmissions might be understood by ET as a taunt. Moreover, the fate of all of mankind cannot be made to be exclusively reliant upon one country's constitution (for example, Holocaust denial is a crime in certain countries in Europe, whereas in the U.S. it is legal pursuant to the First Amendment).

#### 2.6.3 METI is Legal, so Who is to Stop us?

Legislation often takes time to catch up with morality. The abolishment of slavery and universal suffrage are examples. Conceding that METI is legal at the moment, it might be best were it governed (along with post-detection protocols) by regulations at the agency level, laws at the national level, and/or international treaties.

### 2.7 How METI-ists Frame Their Opponents

#### 2.7.1 Opponents are Merely a Few Malcontents

METI-ists are wont to airily dismiss or diminish their opponents. In his New York Times op-ed, Shostak [25] described an anti-METI petition [6] as drafted by "a small consortium of academics." In fact, it was drafted by members of the University of California Berkeley's SETI program, which is the preeminent SETI program in the world, and the recent recipient of most of Yuri Milner's $100 million SETI research grant. The SETI search that the Berkeley group has now commenced will be orders of magnitude more powerful than the aggregate of all prior searches performed by Shostak's SETI Institute (this author freely admits this as a former chairman of the board of the SETI Institute). While drafted at Berkeley, signatories are hardly limited to California academics, as Shostak suggests, and includes such familiar names as Elon Musk, George Dyson, Dan Werthimer, Geoff Marcy, Paul Davies, David Brin, Michael Michaud, James Benford, among others, nor does this petition include such eminent scientists who have elsewhere gone on the record as being opposed to METI as Stephen Hawking [26] Neil DeGrasse Tyson, Sean Carroll [23], Jarod Diamond [27], as well as, before their demise, John Billingham Martin Ryle, and Carl Sagan. SETI founder, Frank Drake, has indicated that METI is a waste of time and money, and has expressed regret about having initiated his 1974 Arecibo transmission to M13.

This author would like to invite the reader to compare for eminence these opponents to METI with the membership of METI International, who intend to commence METI transmissions as soon as they locate the funding to do so [2, 3]. Further, not all opponents to METI have registered their opposition publicly. For example, when Shostak and Vakoch approached the acting CEO of the SI, Edna Devore (the CEO, Tom Pierson, had recently died), with a proposal to commence METI transmissions to the nearest stars, Devore had the good sense to alert the executive board, on which this author served at the time. The executive board passed the proposal along to the full board, which, after due consideration, roundly rejected it.

Seth Shostak is the host of the radio show and podcast, *Big Picture Science*. Each week, he and co-host, Molly Bentley, tackle a single topic in science in a way that is both informative and entertaining. However, their episode devoted to METI, entitled, "How to Talk to Aliens," veers towards propaganda. They neglected to include even a single opponent to METI among their interviewees. Shostak very briefly mentioned that Hawking is opposed, but then Bentley snarkily dismissed this by suggesting that Hawking must think ET is hungry for human flesh. They are not alone in dismissing METI opponents with pejoratives. Musso states that "so many authors are in favor of active SETI," without also mentioning that so many authors are also opposed, or referring to the wide gulf between the two groups in terms of eminence. Musso echoes other METI-ists in referring pejoratively to opponents: "maybe their concerns are actually irrational or childish, as many people in the SETI community think [20]." Not to be outdone, Dick accuses opponents of METI of "xenophobia" who "cower and hide from the stars [3]."

#### 2.7.2 What, Me Worry?

Shostak, in a recent op-ed, dismisses the many of his colleagues who oppose METI of "paranoia based on nothing more than conjecture [25]." With equal evidence (i.e., no evidence whatsoever), Shostak can be accused of *wild euphoria* based on nothing more than conjecture. It is useful to closely parse the exact wording used by METI-ists. With clever twist of tongue, Shuck and Almar [29] admit that METI "is not wholly without risk," thereby leaving the reader with impression that the risk is small, as if they have any idea what the actual risk might be. Similarly, Korbitz, acknowledging a total lack of evidence uses that very lacuna to argue in favor of METI: "Given this vacuum of knowledge, we do not currently have reason to believe that Active SETI is inherently risky [30]." With equal cupidity one might walk in the woods, cloaked in complete mycological ignorance, and commence eating whatever mushroom happens to look delicious. METI opponents have been accused of absolute risk aversion--that even be the risks tiny they would still be opposed to METI. However, opponents of METI have never made any claims about the size or severity of the risk. Probabilistically, all we can know about the risk of a bad outcome from conducting METI is that it lies somewhere between zero and one hundred percent. It is precisely because we have absolutely no idea whether there is a large or a small risk of a bad outcome, or just how bad or wonderful that outcome might be, that we can say nothing whatsoever about the risk profile other than that, along whatever spectrum is chosen, the risk is unknown.

### 2.8 Someone Has to be The First to Transmit

METI-ists argue that one possible solution to Fermi's Paradox is that everyone is listening while no one is transmitting. Maybe it falls upon Earth to get the interstellar conversation going. SETI scientists, including those who promote METI, are in agreement that any civilization we detect will be eons more advanced than ourselves. Are they all too incurious or too timid to transmit? If they are fearful, perhaps they have good reason to be. Why should the very youngest civilization, ourselves, be the first?

### 2.9 Why Wait?

METI-ists are ready to transmit tomorrow if given the keys to a powerful transmitter like Arecibo. Not wishing to seem like petulant children, they allow that, of course, in a perfect world it would be nice to receive general permission to transmit signals by getting, say, a vote from the United Nations Security Council. But that would take too long to achieve [28]. If their cause is truly just, they can argue their case and, eventually, like suffrage or civil rights, they may persuade and prevail. What is the rush? The stars are not going anywhere on human timescales. SETI scientists should have known when they signed up for the mission that this might very well be the quest of generations. None of us may live long enough to witness the day of First Contact. Once we send a METI signal we encumber generations unborn with that decision. We know what it is like to live today with the thoughtlessness of prior generations, such as the destruction of mega-fauna in the Americas, Australia and elsewhere by its first human inhabitants, as one example among very many. Once METI signals are sent they can never be recalled. Post offices and elementary schools may not be named after anyone from the first generation of SETI scientists because they had achieved First Contact through METI. On the other hand, if they desist from METI, at least they will negate the possibility that history would take a very dim view of their METI activity.

## 3. METI METHODOLOGY – WHAT METHODOLOGY?

METI-ists have proposed to send messages to our closest neighboring stars. However, they have made no provision whatsoever for the receipt of ET's return message. For example, were they to send a message to a star at, say, 29 LYs, they should have a plan in place for the receipt of a return message commencing 58 years into the future. If they would transmit from Arecibo today, as Shostak and Vakoch proposed to the board of the SETI Institute, they must also reserve Arecibo, or its equivalent, for that future date. Of course, they have not. In fact, Arecibo will probably be decommissioned long before then. Moreover, to properly cover just that one star 24/7 would require multiple receivers spaced around the globe. Otherwise, ET's return message might wash over Earth undetected simply because the star was beneath a single telescope's horizon at the time. That return message might arrive in 58 years, but it might also take much longer as ET spent time decoding Earth's message and debating within its own society whether and how to respond. Consequently, the METI-ist's receivers should be looking at that star 24/7 for many years after the 58$^{th}$ year, the first possible year of a return message. METI-ists must multiply this procedure for as many stars as they would target. If they wish to allow ten years per star for return message receipt, and target stars at distance intervals of 5 LYs, and if METI-ists had four dedicated radio telescopes spaced at 90 degree intervals around the globe, they would be able to send four messages every five years out to, sequentially, the nearest to the furthest stars, and be ready to receive back messages from these stars at ten year intervals. In a century, they would have sent about 80 messages and been able to retrieve a hypothetical maximum of about 40 return messages. One might still protest the exercise, but at least METI-ists could counter-argue that their methodology is sound. However, having no plan for return message reception, METI-ists court disaster if the recipient be hostile, while not being able to enjoy the beneficence of a return message should ET prove benign.

## 4. CONCLUSIONS

Whenever one hears a "scientist" assert that ET must be altruistic, or that ET surely knows we are here, or that the closet ET civilization is at least $x$ LY away, ask to see the data set on which they base their conclusions. As of today, no such data set exists. In the absence of any evidence whatsoever, whether one believes that the extraterrestrial civilization we might first encounter will be benign, in the fashion of Spielberg's *Close Encounters of the Third Kind*, or *ET*, or malicious, as in Ridley Scott's *Alien*, or robotic, or something else entirely is strictly a matter of one's personal taste. SETI experiments seek to learn what actually resides or lurks out there in the universe. METI plays Russian roulette without even knowing how many bullets are in the chamber. It would be wiser to listen for at least decades if not centuries or longer before we initiate intentional interstellar transmissions, and allow all of mankind a voice in that decision. The power of SETI has grown exponentially with Moore's Law, better instruments, better search strategies, and now thanks to Milner's visionary investment, meaningful funding. The advances are so profound that it is reasonable to say that the SETI of the next 50 years will be many orders of magnitude more powerful than the SETI of the last 50 years. Shostak, perhaps METI's most articulate proponent, knows this and has widely predicted that we will achieve Contact within the next two decades. So why can he and his fellow METI-ists not wait at least until then before initiating transmissions?

A METI experiment based on an actual methodology that includes a plan to receive ET's reply, might leave some to call that method madness, but at least it would qualify as actual science. Sending a message without a practical plan in place to receive a return message, leads to the conclusion that METI transmissions are like a Hail Mary, they have more in common with a faith based religion than with science. METI-ists implicitly believe that ET is omniscient (they know we are here even though our leakage is trivial); all good (ET must be altruistically interested in our welfare); and omnipotent (even though we have made no provision to receive their return message, they will make themselves known to us somehow). It is fair to ask that METI-ists not impose their religion on the rest of us.

* * *